\documentclass[referee,usenatbib]{raa_rb}            

\usepackage{graphicx,times}             
\usepackage{lscape}
\usepackage{amsmath}
\usepackage{textcomp}
\usepackage{multicol}
\usepackage{longtable}
\usepackage{natbib}
\usepackage[colorlinks,linkcolor=cyan,anchorcolor=black,citecolor=cyan]{hyperref}
\newcommand{\Ion}[2]{#1{\,\scriptsize #2}}
\begin{document}

\title{A Sample of E+A Galaxy Candidates in LAMOST Data Released 2\,$^*$
\footnotetext{$*$ Supported by National Key Basic Research Program of China (Grant No. 2014CB845700), the National Natural Science Foundation of China (Grant Nos 11390371, 11403036, 11403059), and the Program for the Outstanding Innovative Teams of Higher Learning Institutions of Shanxi.}
}

\volnopage{Vol.0 (200x) No.0, 000--000}      
\setcounter{page}{1}          

\author{
	Haifeng Yang\inst{1,2,3},
	Ali Luo\inst{1},
	Xiaoyan Chen\inst{1},
	Wen Hou\inst{1,3},
	Jiannan Zhang\inst{1},
	Wei Du\inst{1},
	Jifu Zhang\inst{2},
	Jianghui Cai\inst{2},
	Yanxin Guo\inst{1,3},
	Shuo Zhang\inst{3},
	Yongheng Zhao\inst{1},
	Hong Wu\inst{1},
	Tinggui Wang\inst{4},
	Shiyin Shen\inst{5},
	Ming Yang\inst{1},
	Yong Zhang\inst{6},
	\and Yonghui Hou\inst{6}\\   
     }

   \institute{Key Laboratory of Optical Astronomy, National Astronomical Observatories, Chinese Academy of Sciences, Beijing 100012, China; {~~~~\it E-mail: lal@nao.cas.cn; hfyang@nao.cas.cn}\\
   \and School of Computer Science and Technology, Taiyuan University of Science and Technology, Taiyuan 030024, China;\\
   \and University of Chinese Academy of Sciences, Beijing 100049, China;\\
   \and University of Science and Technology of China, Nanjing 230026, China;
   \and Shanghai Astronomical Observatory, Chinese Academy of Sciences, Beijing 100012, China; 
   \and Nanjing Institute of Astronomical Optics and Technology, National Astronomical Observatories, Chinese Academy of Sciences, Nanjing 210042, China    
   }

   \date{Received~~2009 month day; accepted~~2009~~month day}

\abstract{A sample of 70 E+A galaxies are selected from 37, 206 galaxies in the second data release (DR2) of Large Sky Area Multi--Object Fiber Spectroscopic Telescope (LAMOST) according to the criteria for E+A galaxies defined by \citet{2004A&A...427..125G}, and each of these objects is further visually identified. In this sample, most objects are low redshift E+A galaxies with z $<$ 0.25, and locate in the high latitude sky area with magnitude among 14 to 18~mag in g, r and i bands. A stellar population analysis for the whole sample indicates that the E+A galaxies are characterized by both young and old stellar populations (SPs), and the metal--rich SPs have relatively higher contributions than the metal--poor ones. Additionally, a morphological classification for these objects is performed based on the images taken from the Sloan Digital Sky Survey (SDSS).
}

   \authorrunning{H.-F. Yang, A.-L. Luo, et al.}            
   \titlerunning{A Sample of E+A Galaxy Candidates}  

   \maketitle

%
%
\section{Introduction}           
\label{sect:intro}

E+A galaxies, with strong Balmer absorption lines and little or no nebular emissions in their optical spectra, are widely considered as short--lived but potentially important phase in galaxy evolution \citep{2014RMxAC..44S.179L, 2003ApJ...599..865T, 2004ApJ...609..683T}. The strong Balmer absorption lines (such as H$\delta$) indicate that the stellar population is dominated by A--type stars which must have formed within the last $\sim$ 1 Gyr. However, the lack of optical emission lines (such as H$\alpha$, [\Ion{O}{II}]) implies that star formation is not ongoing in the galaxies. A general interpretation of this spectral feature is that these galaxies are observed in the post--starburst phase (e.g. \citealt{1983ApJ...270....7D,1987MNRAS.229..423C,1999ApJ...518..576P,2003PASJ...55..771G,2007MNRAS.377.1222G}). While there are still some questions about this view, for example, they may also be produced by the abrupt truncation of star formation in a disc, without necessarily requiring a starburst (e.g. \citealt{2004ApJ...601..654S,2005MNRAS.359..949B}). A lot of works were dedicated to investigate the physical mechanisms such as galaxy mergers \citep{1996ApJ...464..641M,2005MNRAS.359..949B}, interaction of physical pairs \citep{2008MNRAS.390..383Y,2011ApJ...729...29M,2005MNRAS.359..949B}, interactions with cluster tidal field \citep{2000ApJ...529..157P,2001ApJ...547L..17B} or intra--cluster medium \citep{1986ApJ...301...57B}. However, the real scenario that can explain such galaxies is still uncertain.

In order to deeply understand the physics of E+A galaxies, enough E+A galaxy samples are needed. However, one of the major difficulties is their rarity \citep{2003PASJ...55..771G}, and only sky surveys can provide the possibility of finding them. Recently, a catalog of E+A galaxies were identified by \cite{2007MNRAS.381..187G} from SDSS (the Sloan Digital Sky Survey) DR5, which includes 564 objects. Seven very local E+A galaxies with 0.0005 $<$ z $<$ 0.01 were discovered by \cite{2012MNRAS.420.2232P} in SDSS DR7. In this paper, we present a sample of 70 local E+A galaxies carefully selected from LAMOST DR2 \citep{2012RAA....12.1197C,2012RAA....12..723Z}. This sample, together with available catalogs, can be used for statistical analyses and follow--up observations in multi--wavelength. 
 
The paper is organized as follows. In Section \ref{sect:Sample}, sample selection and auto--search method are described in details, and how to identify the E+A galaxies are presented. A preliminary discussion about this sample is shown in Section \ref{Discuss}, including the space distribution, stellar population, and image analysis. Finally, a summary is given in Section \ref{summary}. 


\section{The E+A galaxy sample}
\label{sect:Sample}
\subsection{LAMOST dataset}
\label{sect:2.1}

The Large Sky Area Multi--Object Fiber Spectroscopic Telescope (LAMOST, also called the Guo Shou Jing Telescope) is a special reflecting Schmidt telescope with an effective aperture of 3.6 --- 4.9 m and a field of view of 5$^{\circ}$. It is equipped with 4000 fibers, with a spectral resolution of R $\approx$ 1800 and the wavelength ranging from 3800 to 9000 $\mathrm{\AA}$ \citep{2012RAA....12.1197C, 2012RAA....12..723Z}. The LAMOST DR2, based on the pilot survey from 2011 October to 2012 June and  the general survey from 2012 September to 2014 June, contains more than four million spectra with limiting magnitude down to V $\approx$ 19.5 mag \citep{2012RAA....12.1243L}. We take 37, 206 spectra as the initial dataset which are spectroscopically classified as `Galaxy' by LAMOST 1D pipeline \citep{2015arXiv150501570L}. In order to find E+A galaxies as many as possible, the low Signal--to--Noise Ratio(SNR) spectra are also remained. The detailed steps for searching E+A galaxies are described in the next section.

\subsection{Search method}
\label{sect:2.2}
In the searching procedure, the most obvious features: strong H$\delta$, weak or no H$\alpha$, [\Ion{O}{II}], are considered as selection criteria. The final E+A galaxy sample is obtained after the strict identifications of following processes.\\

1. 162 spectra with redshift z $>$ 1.18 are removed since their H$\delta$ lines are not within the spectral wavelength coverage of LAMOST.\\

2. The spectra are shifted to the rest wavelength frame, and the Equivalent Widths (EWs) of H$\delta$, H$\alpha$ and [\Ion{O}{II}] lines are roughly measured in the wavelength window of 4082 --- 4122 \AA, 6553 --- 6573 \AA, and 3717 --- 3737 \AA~\citep{2007MNRAS.381..187G}, respectively. If redshift z $<$ 0.022 where [\Ion{O}{II}] double lines are not in the optical wavelength range of LAMOST, the EW$\rm_{[\Ion{O}{II}]}$ = 99; and if redshift z $>$ 0.37 where H$\alpha$ line is not in this wavelength range, we set the EW$\rm_{H\alpha}$ = 99 \footnote{Emission lines have a negative value throughout this paper, the value 99 indicates that this line should be neglected in selection.}.\\

3. Considering the uncertainty of EWs resulted from measurement errors, noise and estimation inaccuracy caused by using uniform wavelength windows for the various line widths of the spectra with lower SNR, our selection criteria for E+A galaxy candidates are temporarily relaxed by 0.5 \AA~than that used by \cite{2004A&A...427..125G}. Therefore, the galaxies with EW$\rm_{H\delta} >$ 3.5 \AA, EW$\rm_{H\alpha} >$ -3.5 \AA~and EW$\rm_{[\Ion{O}{II}]}>$ -3.5 \AA~will be identified in the next step.\\

4. The spectra satisfying the above conditions are visually inspected one by one in this step. Those spectra with specific redshift in which H$\delta$, H$\alpha$ are contaminated by sky lines or telluric lines are excluded. The spectra with redshift z $\in$ [0.383, 0.445] are also removed from our sample since in this redshift range, H$\delta$ lines of these spectra are just falling in the connection band of blue and red where the features might not be real because of low efficiency of the instrument in this range.\\ 

5. The wavelength window boundary of H$\delta$, H$\alpha$ and [\Ion{O}{II}] are adjusted artificially to keep the completeness of lines. Then, equivalent widths of these three lines are remeasured inside the adjusted wavelength windows.\\

6. The final step of making these E+A galaxy sample is to remove the objects with EW$\rm_{H\delta} <$ 4 \AA, EW$\rm_{H\alpha} <$ -3 \AA~and EW$\rm_{[\Ion{O}{II}]}< $ -2.5 \AA~, which are the same criteria as \citet{2007MNRAS.381..187G} and 70 E+A galaxies are finally identified.\\ 

\subsection{Catalog of 70 local E+A galaxies}
\label{sect:2.3}

Table \ref{table1} in Appendix lists the E+A galaxy sample selected from LAMOST DR2 by the method mentioned above, and Table \ref{table2} lists some corresponding photometric features. This first E+A galaxy sample of LAMOST survey includes 70 objects, in which 56 of them are newly discovered. These E+A galaxies span a redshift range of $0.0034\leq$ z $\leq 0.2541$. 

\begin{figure} 
\centering
\includegraphics[width =\textwidth,bb = 0 0 900 450]{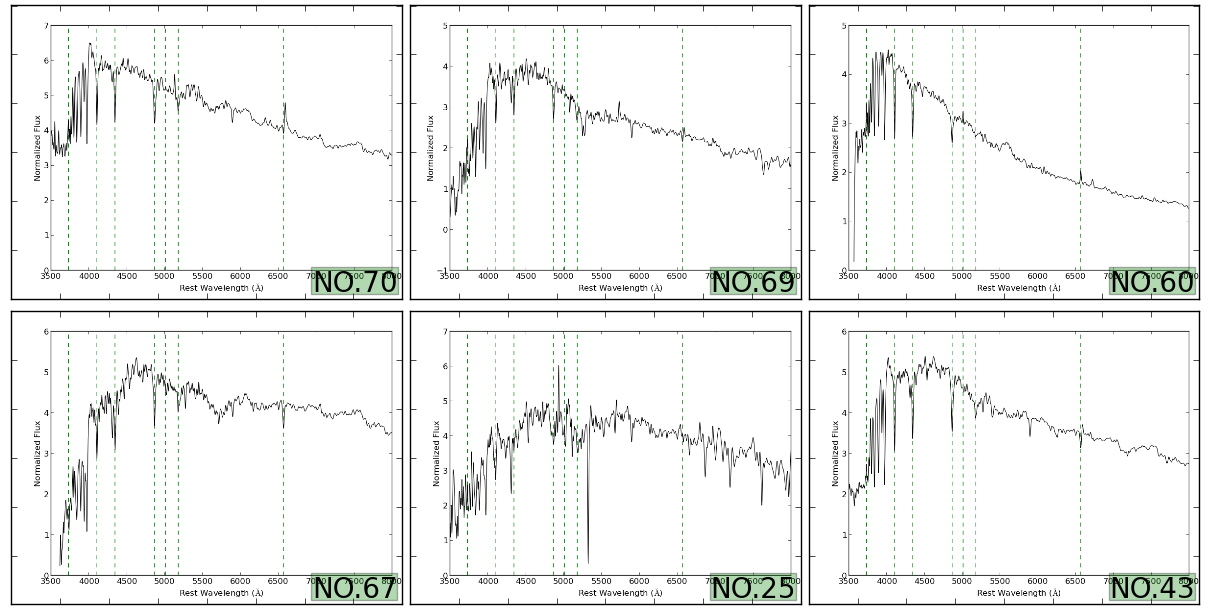}
\caption{Six spectral examples in our E+A galaxy sample are shown in this figure.  In order to show the features clearly, these spectra are smoothed to R$\sim$1000 by Gaussian function and shifted to rest wavelength frame. Some typical lines are marked by green dashed lines and the serial number of each object are marked in the bottom right corner.}{\label{spectra}}
\end{figure}

In Table \ref{table1}, each E+A galaxy is assigned a serial number (shown in column 1) that will be referred to throughout this paper and the signs marked upper right corner indicate that these objects have been published or studied in other literatures. 
The other columns show the basic information of these objects including designation, redshift, RA, DEC, EW$\rm_{H\delta}$, EW$\rm_{H\alpha}$, and EW$\rm_{[\Ion{O}{II}]}$. The redshift, RA, and DEC are copied from LAMOST catalog, and equivalent widths of these lines are measured by the method mentioned in previous section. Fig. \ref{spectra} shows six spectral examples of E+A galaxies, which are corresponding to the morphological classes in Fig. \ref{image} as described in Section 3.3. Table \ref{table2} lists the magnitude of u, g, r, i and z bands, the colors of g-r, r-i, and image class by crossing these targets with available SDSS photometric catalog (discussed in the next section).

\section{Discussions}
\label{Discuss}
\subsection{The distribution feature of the sample}
\label{sect:3.1}
\begin{figure}[h]
\centering
\includegraphics[width = 3.5in,bb = 0 0 600 450]{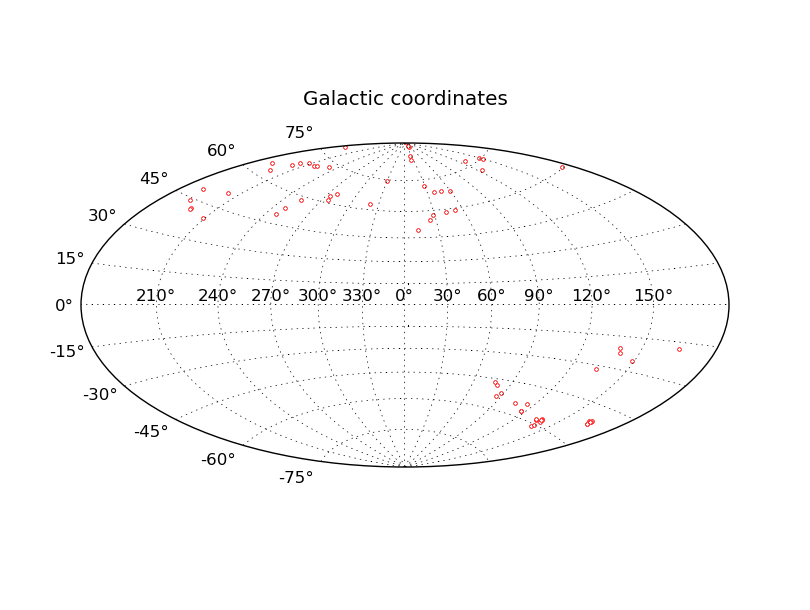}
\caption{This figure shows the spatial distribution of E+A galaxies selected from LAMOST DR2. These objects are marked by red circles in galactic coordinates.}{\label{spatialDistribution}}
\end{figure}

\textit{~~~~Spatial distribution.} The spatial distribution of E+A galaxies in our sample is shown in Fig. \ref{spatialDistribution}. Most  objects locate in high galactic latitude areas, while only a few of them are in the direction of anti--galactic--center. This may not be the true distribution feature of E+A galaxies because of the selection effect of input catalog and relatively small size of our sample. 

\begin{figure}[h]
\begin{minipage}[h]{0.5\linewidth}
\centering
\includegraphics[width = 2in,bb = 0 0 300 300]{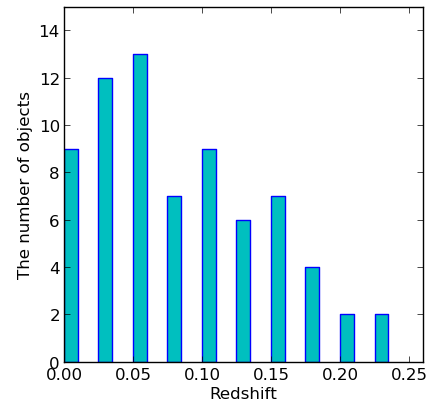}
\caption{The histogram of redshift distribution of the E+A galaxies in our sample. The redshift bin is 0.025. }{\label{zDistribution}}
\end{minipage}
\begin{minipage}[h]{0.5\linewidth}
\centering
\includegraphics[width = 2in,bb = 0 0 300 300]{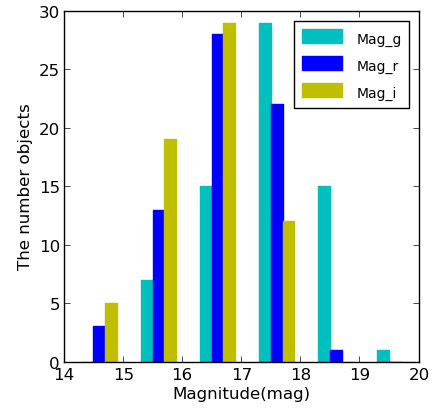}
\caption{The histogram of magnitudes distributions of E+A galaxies. The cyan, blue and yellow bars represent the magnitude distributions of g, r and i bands, respectively.}{\label{magDistribution}}
\end{minipage}
\end{figure}

\textit{Redshift distribution.} The redshift distribution of the E+A galaxies in our sample is shown in Fig. \ref{zDistribution}. The average redshift is 0.0875 and about 65.5\% galaxies have the redshift z $<$ 0.1. The main reason is that LAMOST is probably more suitable for the observations of nearby targets. We did not rule out the presence of a peak at z $\sim$0.04 as described as \cite{2005MNRAS.357..937G}, since the aperture of LAMOST is so  similar with that of SDSS that it will bring an aperture bias for the nearby E+A galaxies due to their large sizes. Nevertheless, these local E+A galaxies, especially for the large apparent size ones, are suitable for more sophisticated researches such as morphology studies, substructure studies, and IFU observations. 

\textit{Magnitude distribution.} Fig. \ref{magDistribution} shows the magnitude distributions of E+A galaxies in our sample. The magnitude of g, r and i bands of these objects are obtained by crossing with SDSS photometric catalog. As seen in this figure, the magnitudes of g, r and i bands of most objects lie between 15 and 18 mag which is the good observational magnitude range of LAMOST. There are also some galaxies with both g and r bands  fainter than 18 mag, however, they are too faint to resolve in the spectra because of their extremely low SNR.

\subsection{Stellar population synthesis}

\begin{figure}[t]
\centering
\includegraphics[width = 5.0in,bb = 0 0 650 300]{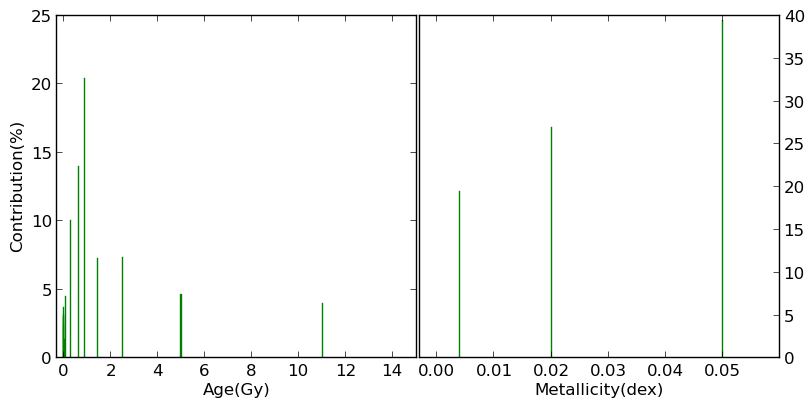}
\caption{The integrated age and metallicity contributions of all spectra in our sample estimated by `STARLIGHT'. The left panel is the age (15 ages : 0.001, 0.003, 0.005, 0.01, 0.025, 0.04, 0.1, 0.3, 0.65, 0.9, 1.4, 2.5, 5, 11 and 13 Gyr) contributions and the right panel is the metallicity (3 metallicities: Z = 0.004, 0.02 and 0.05) ones.}{\label{contribution}}
\end{figure}

Stellar population synthesis are performed to each spectrum by using the stellar population synthesis code `STARLIGHT' \citep{2010A&A...515A.101C}. During the estimation, we restrict the wavelength range as 3700 --- 8500 \AA~(in rest frame), and choose 45 SSPs (Simple Stellar Populations, widely used in stellar population analysis) extracted from BC03 \citep{bruzual2003stellar} as templates. We use the Padova 1994 tracks, the \cite{chabrier2003galactic} initial mass function, and the \cite{calzetti1994dust} reddening law. To understand the contributions of different stellar populations to the whole sample, age and metallicity contributions of all objects are integrated as shown in Fig. \ref{contribution}.

The left panel of Fig. \ref{contribution} indicates the obvious feature of young SPs with age of about 1 Gyr in the E+A galaxy spectra. Meanwhile, the old SPs with age of about 11 Gyr also contribute to these E+A galaxies. It confirms the selection criteria of E+A candidates properly. The right panel of Fig. \ref{contribution} reveals that the metal--rich SPs have relatively higher contributions than metal--poor ones. 

\subsection{Image analysis}
\label{sect:3.3}
\begin{figure*}
\includegraphics[width = \textwidth,bb = 0 0 900 600]{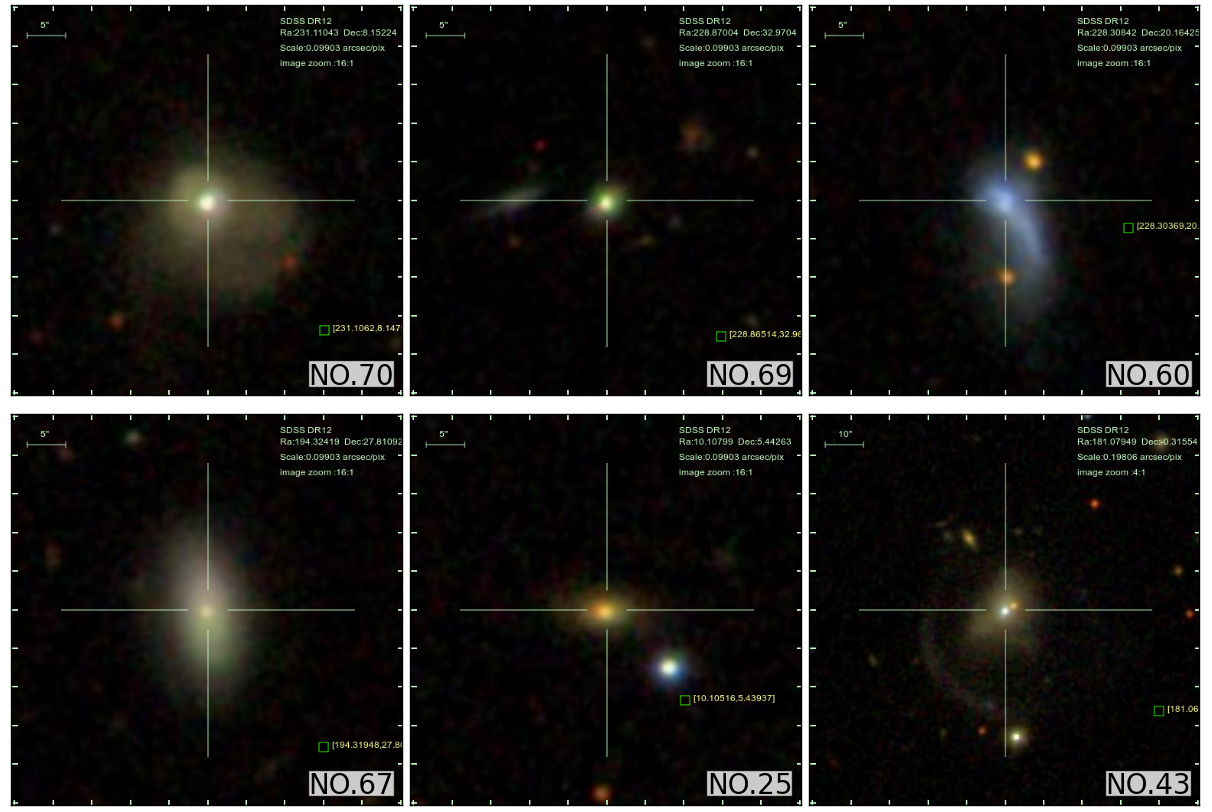}
\caption{The examples of composite images of E+A galaxies. The corresponding spectra are shown in Fig. \ref{spectra}. The images are sorted by image classes shown in Table \ref{table2}. The first column are FOL1 (upper panel, NO.70) and FOL2 (lower panel, NO.67), the second column are FOS1 (upper panel, NO.69) and FOS2 (lower panel, NO.25), the top right corner is EOI (NO.60), and the bottom right corner is SGs (NO.43).}{\label{image}}
\end{figure*}

Many studies indicate that there are different features in bulge, halo and disk of such special galaxies. We have obtained the composite images of the E+A galaxies in our sample by crossing with SDSS photometric catalog. We divided these images into four classes as seen in Table \ref{table2} and Fig. \ref{image}.\\ 
1. Face--On--Small (FOS) are the E+A galaxies facing on to us and projected size $<$ 3".\\ 
2. Face--On--Large (FOL) are the ones facing on to us and projected size $>$ 3". For FOS and FOL E+A galaxies, FOS (FOL)1 indicates the galaxy has a very bright central bulge while FOS (FOL)2 has not.\\ 
3. Edge--On/Irregular(EOI) are the ones edging on to us or irregular galaxies.\\
4. Special galaxies (SGs) are the ones with some special features such as lens--like.  

\section{Summary}
\label{summary}

In this paper, we present a sample of E+A (post--starburst) galaxies based on the selection criteria EW$\rm_{H\delta}>$ 4.0 \AA, EW$\rm_{H\alpha}>$ -3.0 \AA~and EW$\rm_{[\Ion{O}{II}]}>$ -2.5 \AA~, which was described in \cite{2004A&A...427..125G}. At such a threshold, 70 E+A galaxies are identified from the LAMOST DR2. If we use a rigorous the criteria of EW$\rm_{H\delta}>$ 5.0 \AA, only 21 E+A galaxies remained which are specially marked in the Table \ref{table1}. Meanwhile, we exclude dozens of galaxies with 3.0 \AA~ $<$ EW$\rm_{H\delta}<$ 4.0 \AA. These objects may be also important evolutionary stage of E+A galaxies especially for those with good signs of H$\alpha$ and [\Ion{O}{II}] (i.e. they are absorption lines or extremely weak emission lines) in their spectra. In order to synchronize the selection criteria, we do not provide them in this paper.

A preliminary analysis for this sample is carried out, including parameter distribution features, stellar population synthesis, and image classification. This sample, together with available catalogs, can be used for statistical analyses and follow--up observations in various wavelengths.

\begin{acknowledgements}

This work is supported by the National Key Basic Research Program of China (Grant No. 2014CB845700), the National Natural Science Foundation of China (Grant Nos 11390371, 11403036, 11403059), and the Program for the Outstanding Innovative Teams of Higher Learning Institutions of Shanxi. The authors would like to thank the anonymous referee for the many suggestions that have helped improve the manuscript. Besides, the authors are also particularly grateful for the AMD scholarship support.

Guoshoujing Telescope (the Large Sky Area Multi-Object Fiber Spectroscopic Telescope, LAMOST) is a National Major Scientific Project built by the Chinese Academy of Sciences. Funding for the project has been provided by the National Development and Reform Commission. LAMOST is operated and managed by the National Astronomical Observatories, Chinese Academy of Sciences.
\end{acknowledgements}

\appendix                  
\section*{Appendix : Sample List}
\begin{center}\footnotesize \doublerulesep 0.2pt \tabcolsep 10pt
\begin{longtable}{@{}llllllll@{}}
\caption[]{{The E+A galaxies list of LAMOST DR2}\label{table1}}\\
\hline   No. & Designation & Redshift & RA & DEC & EW$\rm_{[\Ion{O}{II}]}$(\AA) & EW$\rm_{H\delta}$(\AA) & EW$\rm_{H\alpha}$(\AA) \\  \endfirsthead

\caption[]{-- continued from previous page}\\  \hline   No. & Designation & Redshift & RA & DEC & EW$\rm_{[\Ion{O}{II}]}$(\AA) & EW$\rm_{H\delta}$(\AA) & EW$\rm_{H\alpha}$(\AA)  \\  \hline  \endhead

\hline \\  \multicolumn{8}{r}{{Continued on next page}} \endfoot
 \endlastfoot
\hline
1 & J111058.63+284038.2 & 0.030500 & 167.744297 & 28.677292 & 23.485 & 4.838 & 1.125\\
2 & J013936.61+294503.6 & 0.069828 & 24.902556 & 29.751000 & -2.159 & 4.458 & 0.580 \\
3$^{+}$ & J105606.79+022818.5 & 0.151924 & 164.028320 & 2.471810 & 4.661 & 6.268 & 0.398 \\
4 & J040244.53+281212.0 & 0.049299 & 60.685553 & 28.203360 & -1.222 & 4.738 & 0.792\\
5$^{*}$ & J131437.60+260726.2 & 0.073523 & 198.656698 & 26.123945 & -2.768 & 4.134 & 0.573\\
6 & J022954.96+031920.2 & 0.157351 & 37.479029 & 3.322285 & -0.690 & 4.098 & -0.840\\
7 & J023255.55+023233.5 & 0.174831 & 38.231477 & 2.542655 & -0.252 & 4.440 & 1.041 \\
8$^{* +}$ & J111742.26+042758.2 & 0.105046 & 169.426090 & 4.466180 & -3.667 & 14.348 & 1.095\\
9 & J023255.55+023233.5 & 0.174737 & 38.231477 & 2.542655 & -0.649 & 4.218 & 0.945\\
10 & J023351.45+023320.2 & 0.137500 & 38.464377 & 2.555622 & -0.898 & 4.030 & 0.751\\
11 & J023255.55+023233.5 & 0.174794 & 38.231477 & 2.542655 & 0.139 & 4.040 & 1.095 \\
12$^{* +}$ & J104349.87+420923.3 & 0.234008 & 160.957814 & 42.156500 & 1.051 & 5.747 & 0.827\\
13 & J125006.26+452930.9 & 0.119162 & 192.526117 & 45.491935 & 1.876 & 4.303 & 0.809\\
14 & J142744.06+345834.5 & 0.127800 & 216.933595 & 34.976265 & 2.554 & 4.625 & 0.442\\
15 & J121837.19+403252.0 & 0.130000 & 184.654973 & 40.547788 & 2.334 & 4.059 & 1.660\\
16 & J121711.03+411740.5 & 0.074964 & 184.295976 & 41.294593 & 3.128 & 4.851 & 1.149\\
17 & J233000.39+031347.6 & 0.058616 & 352.501629 & 3.229915 & -3.296 & 4.487 & 1.058\\
18$^{+}$ & J233226.01+052135.7 & 0.069256 & 353.108400 & 5.359930 & 1.688 & 6.310 & 1.174\\
19 & J024416.24+312200.0 & 0.017100 & 41.067678 & 31.366692 & 2.629 & 4.733 & 1.259\\
20 & J013003.020025652.5 & 0.215250 & 22.512591 & -2.947918 & 1.579 & 4.098 & 0.081\\
21 & J012804.460041257.0 & 0.048006 & 22.018585 & -4.215844 & 5.687 & 4.333 & 1.672\\
22 & J023103.93+014148.1 & 0.021150 & 37.766400 & 1.696709 & 3.212 & 4.125 & 1.377 \\
23 & J225648.87+052901.4 & 0.157710 & 344.203640 & 5.483740 & 2.007 & 4.031 & 0.469\\
24 & J230536.08+053345.4 & 0.051679 & 346.400360 & 5.562620 & 1.364 & 4.107 & 1.221\\
25$^{+}$ & J004025.91+052633.4 & 0.107554 & 10.107990 & 5.442630 & -2.691 & 6.142 & -0.325\\
26$^{+}$ & J004053.76+052115.4 & 0.135811 & 10.224040 & 5.354280 & 3.218 & 5.072 & 0.522 \\
27 & J003409.66+084511.5 & 0.107989 & 8.540290 & 8.753200 & -0.021 & 4.276 & 0.865 \\
28 & J001539.84+064100.6 & 0.069929 & 3.916040 & 6.683520 & 2.991 & 4.054 & 0.794 \\
29 & J012522.270003839.0 & 0.017848 & 21.342811 & -0.644178 & 3.737 & 4.020 & -0.604 \\
30$^{+}$ & J012730.400012319.5 & 0.016193 & 21.876671 & -1.388775 & -1.545 & 5.110 & 1.060 \\
31$^{+}$ & J021415.88+391627.4 & 0.082600 & 33.566198 & 39.274292 & 8.0724 & 7.095 & -2.621\\
32$^{+}$ & J021821.10+363113.9 & 0.037300 & 34.587923 & 36.520552 & 6.2058 & 5.308 & 0.791 \\
33$^{+}$ & J095530.30+255201.8 & 0.184076 & 148.876253 & 25.867171 & -1.734 & 6.939 & 0.546\\
34 & J023115.51+014119.7 & 0.022279 & 37.814643 & 1.688816 & 1.851 & 4.567 & 0.859 \\
35$^{*}$ & J091140.68+274111.2 & 0.157616 & 137.919525 & 27.686467 & -1.870 & 4.383 & 1.770 \\
36 & J091000.34+274635.8 & 0.179827 & 137.501420 & 27.776630 & -0.948 & 4.003 & 0.534 \\
37 & J012904.130034134.8 & 0.203008 & 22.267248 & -3.693003 & 0.371 & 4.236 & 0.400 \\
38$^{+}$ & J012047.570033929.9 & 0.061524 & 20.198220 & -3.658321 & -0.445 & 6.026 & 0.984 \\
39 & J012633.050035540.4 & 0.198778 & 21.637733 & -3.927915 & -0.494 & 4.296 & -1.417\\
40 & J011908.930035346.3 & 0.131500 & 19.787224 & -3.896207 & 0.203 & 4.780 & 0.875\\
41 & J094106.78+344356.7 & 0.049757 & 145.278260 & 34.732440 & 1.197 & 4.367 & 1.038\\
42 & J120329.26+023914.4 & 0.047823 & 180.871929 & 2.654006 & -1.550 & 4.284 & 0.645\\
43$^{* +}$ & J120419.070001855.9 & 0.093631 & 181.079480 & -0.315540 & -0.991 & 6.486 & 1.193\\
44 & J091353.44+185630.4 & 0.032403 & 138.472670 & 18.941793 & 1.8432 & 4.160 & 1.327\\
45$^{* +}$ & J104230.55+003441.9 & 0.100940 & 160.627319 & 0.578313 & -1.482 & 5.968 & -0.451\\
46$^{* +}$ & J105755.26+334041.7 & 0.058130 & 164.480257 & 33.678251 & -2.467 & 5.174 & 1.455\\
47 & J104856.23+295406.7 & 0.197170 & 162.234310 & 29.901880 & -0.087 & 3.945 & 0.726\\
48 & J132556.100004208.6 & 0.055620 & 201.483756 & -0.702390 & 0.407 & 3.952 & 0.997 \\
49$^{*}$ & J112844.18+234053.3 & 0.131520 & 172.184110 & 23.681485 & -0.897 & 4.966 & -0.026\\
50$^{+}$ & J113108.88+230325.7 & 0.031260 & 172.787035 & 23.057152 & 2.351 & 5.095 & 0.780 \\
51 & J132318.78+130630.9 & 0.050370 & 200.828276 & 13.108591 & 0.631 & 3.570 & 1.005 \\
52 & J112512.50+262646.5 & 0.047500 & 171.302120 & 26.446264 & 0.917 & 4.202 & 1.451 \\
53 & J111728.76+275546.4 & 0.046750 & 169.369836 & 27.929575 & -0.709 & 4.009 & 1.073 \\
54 & J132132.00+254816.8 & 0.225000 & 200.383358 & 25.804694 & -0.756 & 4.246 & 1.101 \\
55 & J121351.590033521.5 & 0.080370 & 183.464961 & -3.589329 & -0.772 & 4.925 & 0.468 \\
56 & J151610.04+280412.3 & 0.111600 & 229.041840 & 28.070110 & 2.337 & 4.463 & 1.044 \\
57$^{+}$ & J114347.75+202148.0 & 0.022210 & 175.948994 & 20.363344 & -0.899 & 5.121 & 1.634 \\
58$^{+}$ & J091903.75+322131.9 & 0.116387 & 139.765640 & 32.358870 & -0.783 & 5.616 & 1.069\\
59 & J120605.41+305637.1 & 0.023631 & 181.522582 & 30.943662 & 1.024 & 4.986 & 1.383 \\
60$^{+}$ & J151314.02+200951.2 & 0.038100 & 228.308423 & 20.164249 & -2.271 & 5.845 & -2.045\\
61 & J152006.92+172848.2 & 0.096290 & 230.028860 & 17.480076 & 2.502 & 3.961 & 1.019 \\
62$^{*}$ & J141127.49+243809.5 & 0.053900 & 212.864561 & 24.635985 & -0.756 & 4.648 & 1.016 \\
63 & J142850.16+310329.5 & 0.079250 & 217.209024 & 31.058200 & 0.541 & 4.024 & 0.810 \\
64$^{*}$ & J142848.23+274149.6 & 0.109590 & 217.200991 & 27.697123 & 0.221 & 4.530 & 1.341 \\
65 & J130008.02+274623.6 & 0.028640 & 195.033430 & 27.773226 & -1.976 & 3.793 & 1.399\\
66 & J125137.96+271838.4 & 0.024200 & 192.908190 & 27.310667 & 4.766 & 4.395 & 0.771 \\
67$^{* +}$ & J125717.81+274839.4 & 0.023580 & 194.324220 & 27.810959 & 3.407 & 5.095 & 1.586 \\
68$^{* +}$ & J124534.16+402559.3 & 0.081600 & 191.392350 & 40.433147 & -0.922 & 5.882 & 1.294 \\
69$^{* +}$ & J151528.81+325813.4 & 0.100090 & 228.870056 & 32.970406 & -0.535 & 5.852 & 1.147 \\
70$^{*}$ & J152426.50+080908.2 & 0.087130 & 231.110420 & 8.152294 & -1.267 & 4.842 & 0.960 \\
\hline 
\end{longtable}
\begin{flushleft}
{\sc Notes:}\\
1. `No.' is the serial number for every object and it will be referred to throughout this paper. The items of Designation, Redshift, RA and DEC are obtained from the catalogs of LAMOST DR2.\\
2. The `*' marked upper right corner of serial number indicates that these objects have been published or studied in other literatures.\\
3. The `+' marked upper right corner of serial number indicates that the EW$\rm_{H\delta}>$ 5.0 \AA.
\end{flushleft}
\end{center}

\begin{center} \footnotesize \doublerulesep 0.2pt \tabcolsep 16pt
\begin{longtable}{@{}clllllllc@{}}
\caption[]{{The photometric features of our E+A galaxy sample }\label{table2}}\\
\hline   No. & \multicolumn{5}{c}{Magnitude(mag)} & g-r(dex) & r-i(dex) & Image Class \\
						&	u	&	g	&	r	&	i	&	z	&	&	&		\\
\endfirsthead

\caption[]{-- continued from previous page}\\  \hline    No. & \multicolumn{5}{c}{Magnitude(mag)} & g-r(dex) & r-i(dex) & Image Class  \\
						&	u	&	g	&	r	&	i	&	z	&	&	&	\\  
\hline  \endhead

\hline \\  \multicolumn{9}{r}{{Continued on next page}} \endfoot
 \endlastfoot
\hline
1 & 17.78 & 15.93 & 15.13 & 14.70 & 14.36 & 0.8 & 0.43	& FOL1 	\\
2 & 18.84 & 17.04 & 16.32 & 15.97 & 15.71 & 0.72 & 0.35	& FOL1 	\\
3 & 21.01 & 18.95 & 17.78 & 17.35 & 17.02 & 1.17 & 0.43	& FOL2 	\\
4 & - & - & - & - & - & - & -	& -		\\
5 & 17.62 & 15.79 & 15.04 & 14.71 & 14.46 & 0.75 & 0.33	& FOL1 	\\
6 & 20.17 & 18.47 & 17.39 & 16.95 & 16.60 & 1.08 & 0.44	& FOS2 	\\
7 & 19.73 & 17.93 & 17.03 & 16.65 & 16.41 & 0.9 & 0.38	& FOS1 	\\
8 & 18.43 & 16.52 & 15.50 & 15.08 & 14.70 & 1.02 & 0.42	& FOL2 	\\
9 & 19.73 & 17.93 & 17.03 & 16.65 & 16.41 & 0.9 & 0.38	& FOS1 	\\
10 & 19.72 & 17.95 & 17.09 & 16.69 & 16.39 & 0.86 & 0.4	& SGs \\
11 & 19.73 & 17.93 & 17.03 & 16.65 & 16.41 & 0.9 & 0.38	& FOS1 	\\
12 & 19.44 & 18.04 & 17.18 & 16.93 & 16.75 & 0.86 & 0.25	& SGs 	\\
13 & 19.04 & 17.21 & 16.44 & 16.12 & 15.83 & 0.77 & 0.32	& FOL2 	\\
14 & 19.03 & 17.31 & 16.36 & 15.97 & 15.61 & 0.95 & 0.39	& FOL2 	\\
15 & 20.10 & 18.31 & 17.39 & 17.02 & 16.71 & 0.92 & 0.37	& FOL2 	\\
16 & 18.58 & 16.79 & 16.04 & 15.68 & 15.41 & 0.75 & 0.36	& EOI \\
17 & 17.69 & 16.02 & 15.33 & 15.02 & 14.80 & 0.69 & 0.31	& FOL1 \\
18 & 19.59 & 17.64 & 16.76 & 16.34 & 16.02 & 0.88 & 0.42	& FOS2 	\\
19 & 17.48 & 15.61 & 14.71 & 14.24 & 13.89 & 0.9 & 0.47	& FOL1 	\\
20 & 20.65 & 18.64 & 17.64 & 17.30 & 17.07 & 1 & 0.34	& FOS2 	\\
21 & 18.89 & 17.19 & 16.43 & 16.09 & 15.80 & 0.76 & 0.34	& EOI 	\\
22 & 17.86 & 16.34 & 15.72 & 15.41 & 15.18 & 0.62 & 0.31	& FOL2 	\\
23 & 20.30 & 18.40 & 17.34 & 16.88 & 16.55 & 1.06 & 0.46	& FOL2 	\\
24 & 18.76 & 17.06 & 16.34 & 16.00 & 15.75 & 0.72 & 0.34	& EOI \\
25 & 20.32 & 18.34 & 17.32 & 16.88 & 16.55 & 1.02 & 0.44	& FOS2\\
26 & 19.30 & 17.48 & 16.64 & 16.28 & 15.99 & 0.84 & 0.36	& FOS2\\
27 & 19.55 & 17.31 & 16.26 & 15.77 & 15.38 & 1.05 & 0.49	& FOS2 	\\
28 & 18.83 & 17.20 & 16.53 & 16.21 & 15.95 & 0.67 & 0.32	& SGs \\
29 & 18.90 & 17.39 & 16.67 & 16.32 & 16.09 & 0.72 & 0.35	& FOS2	\\
30 & 18.25 & 16.63 & 15.94 & 15.58 & 15.32 & 0.69 & 0.36	& FOS2	\\
31 & - & - & - & - & - & - & -	& -		\\
32 & - & - & - & - & - & - & -	& -		\\
33 & 19.49 & 17.96 & 17.27 & 16.98 & 16.84 & 0.69 & 0.29	& FOS1 	\\
34 & 18.49 & 16.96 & 16.24 & 15.90 & 15.63 & 0.72 & 0.34	& FOL2 	\\
35 & 18.91 & 17.13 & 16.54 & 16.22 & 15.98 & 0.59 & 0.32	& FOL1 	\\
36 & 20.16 & 18.26 & 17.32 & 16.89 & 16.61 & 0.94 & 0.43	& SGs 	\\
37 & 20.63 & 18.99 & 17.85 & 17.33 & 16.99 & 1.14 & 0.52	& FOS2 	\\
38 & 21.19 & 19.12 & 18.18 & 17.77 & 17.44 & 0.94 & 0.41	& FOS2 	\\
39 & 20.09 & 18.56 & 17.72 & 17.37 & 17.09 & 0.84 & 0.35	& FOS2 	\\
40 & 18.97 & 17.42 & 16.74 & 16.42 & 16.17 & 0.68 & 0.32	& FOS1 	\\
41 & 18.60 & 17.04 & 16.49 & 16.24 & 16.07 & 0.55 & 0.25	& FOS2 	\\
42 & 19.16 & 17.54 & 16.86 & 16.55 & 16.30 & 0.68 & 0.31	& EOI \\
43 & 17.59 & 16.08 & 15.61 & 15.45 & 15.24 & 0.47 & 0.16	& SGs 	\\
44 & 17.07 & 15.41 & 14.79 & 14.48 & 14.26 & 0.62 & 0.31	& FOL1 	\\
45 & 19.22 & 17.44 & 16.80 & 16.47 & 16.24 & 0.64 & 0.33	& FOS2 	\\
46 & 19.49 & 17.71 & 17.09 & 16.78 & 16.57 & 0.62 & 0.31	& FOS2 	\\
47 & 20.28 & 18.60 & 17.65 & 17.30 & 17.06 & 0.95 & 0.35	& FOS2 	\\
48 & 18.40 & 16.77 & 16.13 & 15.84 & 15.61 & 0.64 & 0.29	& FOS1 	\\
49 & 19.00 & 17.21 & 16.38 & 15.97 & 15.62 & 0.83 & 0.41	& SGs \\
50 & 16.95 & 15.25 & 14.48 & 14.10 & 13.83 & 0.77 & 0.38	& FOL2	\\
51 & 19.05 & 17.41 & 16.70 & 16.36 & 16.13 & 0.71 & 0.34	& FOS2 	\\
52 & 18.34 & 16.79 & 16.16 & 15.89 & 15.66 & 0.63 & 0.27	& FOL1 	\\
53 & 19.11 & 17.35 & 16.58 & 16.21 & 15.93 & 0.77 & 0.37	& FOS2 	\\
54 & 19.77 & 18.50 & 17.51 & 17.07 & 16.83 & 0.99 & 0.44	& FOS2 	\\
55 & 18.40 & 16.97 & 16.46 & 16.20 & 16.00 & 0.51 & 0.26	& FOS2 	\\
56 & 20.15 & 18.42 & 17.61 & 17.27 & 16.99 & 0.81 & 0.34	& FOS2 	\\
57 & 17.54 & 16.13 & 15.66 & 15.42 & 15.26 & 0.47 & 0.24	& SGs \\
58 & 19.16 & 17.53 & 16.86 & 16.60 & 16.36 & 0.67 & 0.26	& FOS1 	\\
59 & 19.63 & 18.06 & 17.43 & 17.12 & 16.90 & 0.63 & 0.31	& FOL2 	\\
60 & 16.63 & 15.85 & 15.69 & 15.59 & 15.52 & 0.16 & 0.1	& EOI\\
61 & 18.99 & 17.25 & 16.47 & 16.15 & 15.86 & 0.78 & 0.32	& FOS2	\\
62 & 17.91 & 16.63 & 16.21 & 16.00 & 15.82 & 0.42 & 0.21	& FOL1 	\\
63 & 19.24 & 17.46 & 16.68 & 16.35 & 16.07 & 0.78 & 0.33	& FOS2 	\\
64 & 19.09 & 17.45 & 16.82 & 16.59 & 16.39 & 0.63 & 0.23	& FOL1 	\\
65 & 18.47 & 17.26 &	 16.99 & 16.85 &	16.84 & 0.27 & 0.14	& SGs\\
66 & 17.55 & 16.06 & 15.48 & 15.20 & 14.99 & 0.58 & 0.28	& FOL2 	\\
67 & 17.55 & 16.03 & 15.50 & 15.25 & 15.06 & 0.53 & 0.25	& FOL2 	\\
68 & 18.21 & 16.56 & 15.91 & 15.61 & 15.39 & 0.65 & 0.3	& FOL1	\\
69 & 19.66 & 17.98 & 17.35 & 17.05 & 16.83 & 0.63 & 0.3	& FOS1	\\
70 & 17.31 & 15.86 & 15.30 & 15.02 & 14.81 & 0.56 & 0.28	& FOL1 \\
\hline
\end{longtable}
\begin{flushleft}
{\sc Notes:}\\
1. `No.' is the serial number for every object and it will be referred to throughout this paper. The magnitudes of u, g, r, i, z, and color of g-r and r-i are obtained from the photometric catalog of SDSS.\\
2. We assign `-' for three objects with no photometric information in SDSS catalog.\\
3. The Image Class is visually assigned for each object based on the Section 3.3.
\end{flushleft}
\end{center}

\bibliography{ms2225}

\label{lastpage}

\end{document}